\documentclass[aps,prb,floatfix,twocolumn,showpacs]{revtex4}
\usepackage{amsmath,amssymb,graphicx,bm}
\usepackage{xcolor}

\IfFileExists{microtype.sty}{\usepackage{microtype}}{\relax}
\usepackage{hyperref}
\hypersetup{
 pdfpagemode=UseNone,
 pdfstartview=FitH
}

\def\veck{\mathbf k}
\def\vecq{\mathbf q}
\def\vecv{\mathbf v}

\def\GGq#1{G^{R}_{\veck#1_{+}}(E_{+}) G^{A}_{\veck#1_{-}}(E_{-})}
\def\rmi{{\rm i}}
\def\rme{{\rm e}}
\def\O{\mathop{\rm O}\nolimits}

\definecolor{darkred}{rgb}{0.6,0.0,0.0}

\begin{document}
\title{Quantum diffusion in a random potential: A consistent perturbation theory}

\author{V.  Jani\v{s}}
\affiliation{Institute of Physics, The Czech Academy of Sciences, Na Slovance 2, CZ-18221 Praha 8, Czech Republic}
 
\email{janis@fzu.cz}

\author{J. Koloren\v{c}}
\affiliation{Institute of Physics, The Czech Academy of Sciences, Na Slovance 2, CZ-18221 Praha 8, Czech Republic}

\date{\today} 
\begin{abstract}
We scrutinize the diagrammatic perturbation theory of noninteracting electrons in a random potential with the aim to accomplish a consistent comprehensive theory of quantum diffusion. Ward identity between the one-electron self-energy and the two-particle irreducible vertex is generally not guaranteed in the perturbation theory with only elastic scatterings. We show how the Ward identity can be established in practical approximations and how the functions from the perturbation expansion should be used to obtain a fully consistent conserving theory. We derive the low-energy asymptotics of the conserving full two-particle vertex from which we find an exact representation of the diffusion pole and of the static diffusion constant in terms of Green functions of the perturbation expansion. We illustrate the construction on the leading vertex corrections to the mean-field diffusion due to maximally-crossed diagrams responsible for weak localization.       
\end{abstract}
\maketitle

\section{Introduction}

Diffusion of particles drifted in inhomogeneous materials by external forces manifests itself on large distances and hence it is a macroscopic phenomenon. That is why it would be treated mostly classically by means of the diffusion equation or variants of the Boltzmann transport equation.\cite{Reichl:1980aa}  Observed deviations from the behavior predicted by the Boltzmann equation are of quantum origin. To explain them we have to develop a microscopic quantum theory of diffusion where elementary objects are waves. Their diffusion is caused by wave scatterings on irregularly distributed impurities in otherwise regular environment. Possible interference of waves makes the qualitative difference between the classical and quantum diffusion.  A theory of multiple scatterings of waves on a random potential was introduced by Lax.\cite{Lax:1951aa,Lax:1952aa} The theoretical framework for the microscopic description of quantum transport and diffusion was set by the linear-response theory of Kubo.\cite{Kubo:1957aa} The interest in microscopic diffusion of quantum particles in disordered systems was boosted by P. W. Anderson  who discovered that charge diffusion due to multiple electron scatterings on randomly distributed impurities can vanish if the strength of the random potential is sufficiently large.\cite{Anderson58} The full understanding of the Anderson localization transition, that is, the transition from diffusive to insulating phase, is still elusive.\cite{IJMPB10,Evers08}

There are two major streams in the study of suppression and eventual vanishing of diffusion: finite-cluster numerical simulations\cite{Kramer93,Markos06} and (semi-)analytic approaches in the thermodynamic limit.\cite{IJMPB10} Unfortunately, agreement between the results and conclusions of the numerical simulations and the analytic theories are not yet fully satisfactory. Each approach has its advantages and drawbacks. The numerical simulations give rather accurate values of the boundary of the localized phase and of the critical exponents, their results need, however, to be appropriately scaled to get rid of finite-size effects. The theoretical approaches in the thermodynamic limit directly access thermodynamically relevant quantities, but suffer from the fact that only homogeneous quantities are available and some of the relevant questions cannot be addressed.

So far, most of the theoretical approaches to quantum transport work well either in the metallic regime, far from the localization transition, or just in the critical region of the transition. Perturbation, diagrammatic expansions in powers of the random potential formulated in the language of the averaged Green functions have been used to describe the metallic phase.\cite{Elliot74} Transformations to effective solvable models then address long-range fluctuations and the critical behavior of correlation and response functions at the Anderson localization transition in the strong-disorder limit.\cite{Mirlin00} Neither of these approaches can smoothly be extended to the opposite limit. The diagrammatic approach works well for one-electron functions and for equilibrium thermodynamic properties but even its self-consistent extension fails to incorporate adequately vertex corrections to the semiclassical Drude expression for the electrical conductivity.\cite{Velicky:1969aa,Khurana:1990aa} The latter construction does not cover adequately the non-universal and non-critical properties of the disordered systems away from the localization transition.\cite{Evers08} It is the diagrammatic approach that is fully microscopic and has the potential to offer a fully consistent first-principles theory of quantum diffusion. 

There are presently two diagrammatic approaches to the microscopic description of quantum diffusion aiming at covering both metallic and insulating phases. It is the self-consistent theory of Anderson localization of Vollhardt and W\"olfle\cite{Vollhardt80a,Vollhardt80b,Vollhardt92} and the parquet approach of the present authors.\cite{Janis05a, Janis05b}  The common quality of both approaches is a two-particle self-consistency used to describe the Anderson localization transition.\cite{Janis11} They differ, however, in the way this self-consistency is used. The former approach uses the framework delimited by the Ward identity and introduces a self-consistent equation for the diffusion constant governing the long-range fluctuations of the singular part of the electron-hole correlation function. Macroscopic conservation laws are obeyed in this approach. A systematic and controllable way of improving the self-consistent equation for the diffusion constant is, however, missing. The latter approach allows one to develop controllable approximations for the two-particle vertices including two-particle self-consistency, but it is unable to reconcile the calculated vertex with the Ward identity and macroscopic conservation laws.

The aim of this paper is to present a systematic and analytically controllable diagrammatic description of diffusive transport in models of non-interacting electrons scattered on randomly distributed impurities. The non-local two-particle scattering processes calculated in the full two-particle state space lead to vertices and Green functions that generically do not comply with the Ward identity and hence do not constitute a conserving theory.\cite{Janis09} To arrive at a conserving approximation, the two-particle irreducible vertex from the diagrammatic construction is only an auxiliary function from which the physical one is appropriately constructed. We show how to make diagrammatic theories conserving and how to restore the Ward identity in approximate treatments. The physical irreducible vertex obeying the Ward identity is then used in a Bethe-Salpeter equation to determine the physical two-particle functions from which all measurable quantities are determined. Thereby, the construction of the two-particle functions becomes fully consistent and the relevant macroscopic thermodynamic relations in the diffusive regime hold.\cite{Janis03} We further derive an exact form of the diffusion pole and of the diffusion constant and apply them to derive leading non-local corrections to the mean-field diffusion constant.       

The paper is organized as follows.  We introduce  the model, notation and basic relations in Sec. II. The way the dynamical Ward identity is established in perturbation theory is presented in Sec. III. We derive in Sec. IV  a low-energy singularity in the full two-particle vertex  and with the aid of the dynamical Ward identity also the diffusion pole in the  electron-hole correlation function. The exact form of the static diffusion constant and its approximate form including maximally crossed diagrams are presented in Sec. V. Section VI brings concluding discussion.       


\section{Definitions, notation and basic relations}

We will consider a noninteracting lattice electron gas scattered on random impurities described by the Anderson tight-binding Hamiltonian
\begin{eqnarray}\label{eq:AD_hamiltonian}
\widehat{H} &=&\sum_{<ij>}t_{ij}\widehat{c}_{i}^{\dagger}
\widehat{c}_{j}+\sum_iV_i \widehat{c}_{i }^{\dagger } \widehat{c}_{i}\ ,
\end{eqnarray}
where $V_i$ is a local, site-independent random potential. We assume validity of the ergodic theorem leading to the existence of the thermodynamic limit, in which all quantities are homogeneous, configurationally averaged. We use the standard diagrammatic representation of the scattering processes.\cite{Elliot74} The state space is a Hilbert space spanned over the Bloch waves $\psi_{\mathbf{k}}(E)$ with energy $E$ and wave vectors $\mathbf{k}$ from the first Brillouin zone.

The fundamental function carrying all information about the linear
response of the disordered electron system is the two-particle Green function
\begin{equation}
  \label{eq:av_2PP}
  G^{(2)}_{ij,kl}(z_1,z_2)=
    \left\langle\left[z_1\widehat{1}-\widehat{H}\right]^{-1}_{ij}
    \left[z_2\widehat{1}-\widehat{H}\right]^{-1}_{kl}
    \right\rangle_{\rm av},
\end{equation}
where the brackets $\langle\:\rangle_{\rm av}$ indicate the configurational averaging. The indices $i,j,\ldots$ correspond to lattice sites
with positions $\mathbf R_i,\mathbf R_j$\ldots. Since the averaging recovers 
homogeneity we use the Fourier transform to the momentum space to label the
averaged Green functions. 


Only real energies are relevant for physical measurable quantities. Nevertheless, we have to
distinguish the way the real energy is reached from the complex
plane to obtain unambiguous results. Energies with an infinitesimal
imaginary part $\eta$ will be tacitly assumed in all expressions, and
superscript $R$/$A$ will correspond to positive/negative $\eta$. We
will use the following notation for the Fourier transform of
the two-particle Green function ($\hbar =1$) 
\begin{multline}\label{eq:2P_momentum}
  G^{RA}_{{\bf k}_+{\bf k}'_+}(E;\omega,{\bf q})
   =  \frac1N\sum_{ijkl} 
  \rme^{-\rmi({\bf k} + {\bf q}/2)\cdot{\bf R}_i}
  \\
 \times \rme^{\rmi({\bf k}'+{\bf q}/2)\cdot{\bf R}_j}
  \rme^{-\rmi({\bf k}'-{\bf q}/2)\cdot{\bf R}_k}
  \rme^{\rmi({\bf k} - {\bf q}/2)\cdot{\bf R}_l}\\[-.7em]
\times G^{(2)}_{ij,kl}
(E + \omega/2 + \rmi0^+,E - \omega/2 - \rmi0^+)\,.
\end{multline}
The independent incoming and outgoing variables of the electron are $\veck_+=\veck+\vecq/2$ and $\veck'_+=\veck'+\vecq/2$, and $\vecq$ is the difference between the incoming (outgoing) momentum of the electron and the hole. 

The principal function of interest in studying quantum diffusion is the electron-hole correlation function. It displays a low-energy singularity, a diffusion pole controlling the long-range electron diffusion\cite{Janis03}
 \begin{align}
 \label{eq:Phi}
 \Phi(E;\omega,\vecq)&=\frac1{N^2}\sum_{\veck\veck'}
 G^{RA}_{\veck\veck'}(E;\omega,\vecq)
 \nonumber \\
 &\xrightarrow[\omega\to0,q\to 0]{} \frac{2\pi n_{F}}{-i\omega + D(\omega)q^{2}}\,,
 \end{align} 
 where we denoted $n_{F}$ the density of states at the Fermi energy $E$ and $D(\omega)$ the dynamical diffusion constant. The electron-hole correlation function contains, however, only a reduced information and is not an integral part of the perturbation theory. It is the two-particle Green function $G^{RA}_{{\bf k}{\bf k}'}(E;\omega,{\bf q})$ that comes out of the two-particle perturbation (diagrammatic) theory. It is the primary objective of the perturbation theory of quantum diffusion to identify the  scattering events contributing to $G^{RA}_{{\bf k}{\bf k}'}(E;\omega,{\bf q})$ relevant for recovering the singular low-energy asymptotics of the electron-hole correlation function $\Phi(E;\omega,\vecq)$.    

To determine the two-particle Green function one has to sum contributions from multiple scatterings to the one- and two-particle functions. To do so effectively one introduces the concept of irreducibility that is best defined diagrammatically. The one-particle irreducibility is uniquely defined and means that cutting a single one-electron propagator does not disconnect the diagram. The contribution from all one-particle irreducible diagrams is contained in the self-energy $\Sigma^{R/A}_{\veck}(E)$ that renormalizes the one-electron propagator via the Dyson equation. The two-particle irreducibility is, however, ambiguous and demands distinction between local and nonlocal scatterings.\cite{Janis01b,Janis05a,Janis05b} Here we do not go into these subtle details of the two-particle perturbation theory. We introduce only the vertex functions that are necessary for the determination of the electron-hole correlation function. We first  single out the uncorrelated propagation from the two-particle Green function 
$G^{RA}_{{\bf k}{\bf k}'}(E;\omega,{\bf q})$ and introduce a two-particle vertex 
$\Gamma^{RA}_{{\bf k}{\bf k}'}(E;\omega,{\bf q})$,
\begin{multline}
G^{RA}_{{\bf k}_+{\bf k}'_+}(E;\omega,{\bf q})=
G^R_{\veck_+}(E_+)G^A_{\veck_-}(E_-)
\bigl[N\delta_{\veck\veck'}
\\
+\Gamma^{RA}_{{\bf k_{+}}{\bf k}'_{+}}(E;\omega,{\bf q})
  G^R_{\veck'_+}(E_+)G^A_{\veck'_-}(E_-)\bigr]\,,
\end{multline}
where we shortened the notation with the aid of definitions $\veck_\pm=\veck\pm\vecq/2$ and $E_\pm=E\pm\omega/2$. The vertex $\Gamma^{RA}_{{\bf k_{+}}{\bf k_{+}}'}(E;\omega,{\bf q})$ contains only the correlated propagation. It can further be split into reducible and irreducible contributions from simultaneous scatterings of pairs of particles. The electron-hole irreducibility is introduced in the diagrammatic representation so that a two-particle diagram cannot be disconnected by cutting just a single pair of electron-hole (antiparallel)  lines. The full vertex is a sum of the irreducible and reducible contributions  that can be represented by a Bethe-Salpeter equation      
\begin{multline} \label{eq:BS-fundamental}
\Gamma^{RA}_{\mathbf{k}_{+}\mathbf{k}'_{+}}(E;\omega,\mathbf{q})
\\
 = L^{RA}_{\mathbf{k}_{+}\mathbf{k}'_{+}}(E;\omega,\mathbf{q})
 + \frac 1N \sum_{\mathbf{k}''} L^{RA}_{\mathbf{k}_{+}\mathbf{k}''_{+}}(E;\omega,\mathbf{q}) 
 \\
 \times  G^{R}_{\mathbf{k}''_{+}}(E_{+}) G^{A}_{\mathbf{k}''_{-}}( E_{-}) \Gamma^{RA}_{\mathbf{k}''_{+}\mathbf{k}'_{+}}(E;\omega,\mathbf{q}) \ .
\end{multline}
 Function  $L^{RA}_{{\bf k}{\bf k}'}(E,\omega;{\bf q})$ is the electron-hole
irreducible vertex containing scatterings that are connected by more than one pair of electron-hole lines. This two-particle function is not independent of the one-particle self-energy.
  Vollhardt and W\"olfle \cite{Vollhardt80b} proved that if a Ward identity    
\begin{multline}\label{eq:WI-VW}
\Delta \Sigma_{\mathbf{k}}^{RA}(E;\omega,\mathbf{q})  \\ = \frac 1{N}\sum_{\mathbf{k}'} L^{RA}_{\mathbf{k}_{+},\mathbf{k}'_{+}}(E, \omega;\mathbf{q}) \Delta G_{\mathbf{k}'}^{RA}(E;\omega,\mathbf{q}) 
\end{multline}
holds then macroscopic conservation laws are obeyed.  
Here we introduced 
$
\Delta G_{\mathbf{k}}^{RA}(E;\omega,\mathbf{q})  = G^{R}_{\mathbf{k}_{+}}(E_{+}) - G^{A}_{\mathbf{k}_{-}}(E_{-}) $
and
$\Delta \Sigma_{\mathbf{k}}^{RA}(E;\omega,\mathbf{q})  = \Sigma^{R}_{\mathbf{k}_{+}}(E_{+}) - \Sigma^{A}_{\mathbf{k}_{-}}(E_{-}) $. 

Perturbation theory of quantum diffusion has to address simultaneously both one- and two-particle irreducible functions, $\Sigma^{R/A}$ and $\Lambda^{RA}$. We demonstrated in our earlier publications  that it is  generally impossible to guarantee validity of the dynamical Ward identity in the perturbation expansion if we go beyond the local mean-field approximation of the irreducible functions, since it is in conflict with causality of the self-energy.\cite{Janis01b,Janis05a,Janis05b} It means that Eq.~\eqref{eq:WI-VW} is generally not obeyed in its full extent. A digression from the Ward identity is a universal feature of any expansion around the local mean-field solution. The problem is that the Ward identity dynamically restricts the correlated movement of two particles in an uncontrollable way.\cite{Janis04c} Ward identity in noninteracting system is not a consequence of a microscopic symmetry transformation, since the one- and two-particle functions are not dynamically coupled. There is hence no microscopic law forcing the Ward identity in random noninteracting systems be strictly obeyed in virtual processes represented by individual scattering events.  To make the perturbation expansion accomplishable we then relax the full dynamical Ward identity when summing diagrammatic contributions to the vertex functions. We denote $\Lambda^{RA}$  the irreducible vertex that is obtained from the perturbation expansion and $\widetilde{\Gamma}^{RA}$ the full two-particle vertex resulting from the corresponding Bethe-Salpeter equation. We must strictly distinguish the vertex functions used in the two-particle perturbation theory from the thermodynamically consistent vertices that do not contradict macroscopic conservation laws.

\section{Establishing Ward identity in the perturbation theory}

The Ward identity is a microscopic condition for making the theory conserving. Quantum theory contains virtual processes represented by Feynman diagrams for which macroscopic conservation laws need not hold unless forced by microscopic (local) symmetry transformations. There are no such symmetries in noninteracting systems. It is only important and necessary that the conservation laws are restored in measurable quantities. To achieve this we  reconcile the one-electron Green functions with the two-particle vertex via the Ward identity in the best possible way. The full dynamical Ward identity can neither be used to determine the one-particle self-energy from the two-particle irreducible vertex nor vice versa, since the vertex contains more information than the self-energy. The Ward identity generally serves only as a consistency check and a guarantee that the macroscopic conservation laws are obeyed. Ward identity~\eqref{eq:WI-VW} for $\omega=0$ and $q=0$ can nevertheless be used to determine the imaginary part of the self-energy from the electron-hole irreducible vertex      
\begin{subequations}\label{eq:KK-relations}
\begin{align}\label{eq:KK-imaginary}
\Im \Sigma^{R}_{\mathbf{k}}(E) & = \frac 1N\sum_{\mathbf{k}'}\Lambda^{RA}_{\mathbf{k}\mathbf{k}'}(E;0,\mathbf{0}) \Im G^{R}_{\mathbf{k}'}(E) \,,
\end{align}
since both sides of this identity contain the same number of degrees of freedom and the equation can consistently be resolved for each energy $E$ and momentum $\veck$. The corresponding real part is then found from the Kramers-Kronig relation
\begin{align}\label{eq:KK-real}
\Re \Sigma^{R}_{\mathbf{k}}(E) & = \Sigma_{\infty} + P\int_{-\infty}^{\infty} \frac{d\omega}{\pi} \frac{\Im\Sigma^{R}_{\mathbf{k}}(\omega)}{\omega - E}
\end{align}
\end{subequations}
that ensures analyticity and causality of the self-energy in the plane of complex energies beyond the real axis. 

Using the one-electron Green functions with the self-energy from Eqs.~\eqref{eq:KK-relations} in the equations for the two-particle vertices is not enough to guarantee validity of the full Ward identity. The Ward identity for nonzero frequencies and transfer momenta must be obeyed to guarantee macroscopic conservation laws. The non-local two-particle irreducible vertex $\Lambda_{\mathbf{k}\mathbf{k}'}(E;\omega,\mathbf{q})$ does not obey the full Ward identity and hence it has to be appropriately modified on a subspace on which its action is determined by the Ward identity. We introduce a new function measuring the deviation of the given vertex $\Lambda^{RA}$ from the Ward identity 
\begin{multline}\label{eq:R-def}
R_{\mathbf{k}}(E;\omega,\mathbf{q})  = \frac 1N\sum_{\mathbf{k}'}\Lambda^{RA}_{\mathbf{k}_{+}\mathbf{k}'_{+}}(E;\omega,\mathbf{q}) \Delta G_{\mathbf{k}'}(E;\omega,\mathbf{q})
\\
 - \Delta \Sigma_{\mathbf{k}}(E;\omega,\mathbf{q}) \ .
\end{multline}
This function vanishes in the metallic phase for $\omega=0$ and $q=0$ due to the definition of the self-energy, Eq.~\eqref{eq:KK-relations}, used in the one-electron Green functions determining the vertex $\Lambda^{RA}$. With the aid of function $R_{\mathbf{k}}$ we construct a new electron-hole irreducible vertex
\begin{multline}\label{eq:L-Lambda}
L^{RA}_{\mathbf{k}_{+}\mathbf{k}'_{+}}(E;\omega,\mathbf{q}) = \Lambda^{RA}_{\mathbf{k}_{+}\mathbf{k}'_{+}}(E;\omega,\mathbf{q})  - \frac 1{\langle\Delta G(E;\omega,\mathbf{q})^{2}\rangle}
\\
\times \left[\Delta G_{\mathbf{k}}(E;\omega,\mathbf{q}) R_{\mathbf{k}'}(E;\omega,\mathbf{q}) + R_{\mathbf{k}}(E;\omega,\mathbf{q}) \Delta G_{\mathbf{k}'}(E;\omega,\mathbf{q}) \phantom{\frac 12}
\right. \\ \left. 
 - \frac{\Delta G_{\mathbf{k}}(E;\omega,\mathbf{q}) \Delta G_{\mathbf{k}'}(E;\omega,\mathbf{q})}{\left\langle\Delta G(E;\omega,\mathbf{q})^{2}\right\rangle}
 \left\langle R(E;\omega,\mathbf{q}) \Delta G(E;\omega,\mathbf{q})\right\rangle\right]
\end{multline}
that now obeys Eq.~\eqref{eq:WI-VW}. We abbreviated $\langle\Delta G(E;\omega,\mathbf{q})^{2}\rangle = N^{-1}\sum_{\mathbf{k}} \Delta G_{\mathbf{k}}(E;\omega,\mathbf{q})^{2}$ and $\left\langle R(E;\omega,\mathbf{q}) \Delta G(E;\omega,\mathbf{q})\right\rangle = N^{-1}\sum_{\mathbf{k}}R_{\mathbf{k}}(E;\omega,\mathbf{q}) \Delta G_{\mathbf{k}}(E;\omega,\mathbf{q})$. Function $L^{RA}$ is the desired physical irreducible vertex to be  used in Eq.~\eqref{eq:BS-fundamental} to determine the physical value $\Gamma^{RA}$ from which all relevant  macroscopic quantities will be calculated. 

Since the Ward identity cannot be fully obeyed in the diagrammatic perturbation theory, vertex functions $\widetilde{\Gamma}$ and $\Gamma$ differ and are equal only when the difference function $R_{\mathbf{k}}(E;\omega,\mathbf{q})$ vanishes. It happens for $\omega=0$ and $q=0$, that is,
\begin{equation}
\widetilde{\Gamma}^{RA}_{\mathbf{k}\mathbf{k}'}(E;0,\mathbf{0}) = \Gamma^{RA}_{\mathbf{k}\mathbf{k}'}(E;0,\mathbf{0})  \ ,
\end{equation}
assuming the self-energy is determined from the two-particle vertex via Eqs.~\eqref{eq:KK-relations}. It means that the vertex function $\Lambda^{RA}_{\veck\veck'}(E;\omega,\mathbf{q})$  is directly related to the measurable macroscopic quantities only for $\omega=0$, $\vecq =\mathbf{0}$.    

\section{Singular behavior of two-particle functions}

\subsection{Low-energy asymptotics of the full two-particle vertex}

Validity of  the dynamical Ward identity, Eq.~\eqref{eq:WI-VW},  is essential for the existence of a low-energy singularity in the two-particle vertex.   Singularity of vertex $\Gamma^{RA}_{\mathbf{k}_{+}\mathbf{k}'_{+}}(E;\omega,\mathbf{q})$ has, however, a more complicated structure than the diffusion pole in the electron-hole correlation function in Eq.~\eqref{eq:Phi}, since the fermionic momenta $\veck,\veck'$ remain relevant and we cannot disregard them.  

To derive the exact singular asymptotics of the full two-particle vertex  in the metallic phase we assume analyticity of the irreducible vertex $L^{RA}_{\mathbf{k}_{+}\mathbf{k}'_{+}}(E;\omega,\mathbf{q})$ in the low-energy limit  $\omega\to 0$  and $q\to 0$ so that an expansion in powers of $\omega$  and $\vecq$ exists. As a first step we rewrite the product of two Green functions as a fraction containing differences of one-particle functions,
\begin{multline}\label{eq:GG-DeltaG}
G_{\veck_{+}}^{R}(E_{+}) G_{\veck_{-} }^{A}(E _{-}) 
\\
=
\frac{\Delta G_{\veck}(E;\omega,\vecq)}{\Delta
  \Sigma_{\veck}(E;\omega,\vecq) - \omega + \Delta
  \epsilon_{\veck}(\vecq)}\,.
\end{multline}
We next use a small-momentum expansion 
\begin{multline}
\label{eq:qexpansion-generic}
f_{\veck\pm\vecq/2}(z)=f_{\veck}(z)
\pm\frac12 q (\hat\vecq\cdot\nabla_{\veck})f_{\veck}(z)
\\
+\frac18 q^2 (\hat\vecq\cdot\nabla_{\veck})
  (\hat\vecq\cdot\nabla_{\veck})f_{\veck}(z)
+\O(q^3)
\end{multline}
to get
\begin{subequations}
\label{eq:qexpansion-components}
\begin{multline}
G_{\veck_{+}}^{R}(E_{+}) G_{\veck_{-} }^{A}(E _{-}) = \frac{\Delta G_\veck(E;\omega,\vecq)}{\Delta\Sigma_\veck(E;\omega,\vecq)} \biggl[1 \\
+ \frac{\omega -\Delta\epsilon_\veck(\vecq) }{\Delta\Sigma_\veck(E;0,\vecq)} - \frac{\Delta\epsilon_\veck(\vecq)^{2} }{\Delta\Sigma_\veck(E;0,\mathbf{0})^{2}} \biggr],
\end{multline}
where
\begin{equation}
\Delta\epsilon_\veck(\vecq)=q(\hat\vecq\cdot\vecv_\veck)+\O(q^3)\,,
\end{equation}
\begin{multline}
\Delta\Sigma_\veck(E;\omega,\vecq)
=2i\Im\Sigma^{R}_\veck(E) + \omega \Re\dot{\Sigma}^{R}\\
+ q(\hat\vecq\cdot\nabla_\veck)\Re\Sigma_\veck^R+\O(q^2)\,,
\end{multline}
and
\begin{multline}
\Delta G_\veck(E;\omega,\vecq)
= 2i\Im G^{R}_\veck(E) + \omega\Re\dot{G}^{R}\\
 + q\bigl[(\Re G_\veck^R)^2-(\Im G_\veck^R)^2\bigr] 
    \vecq\cdot(\vecv_\veck+\nabla_\veck\Re\Sigma_\veck^R)
\\
  -2\,\Re G_\veck^R\,\Im G_\veck^R\, 
  (\hat\vecq\cdot\nabla_\veck)\Im\Sigma_\veck^R
  +\O(q^2)\,.
\end{multline}
\end{subequations}
The dot indicates the derivative with respect to $\omega$. We further introduce two new vertex functions
\begin{align}
\mathcal{L}_{\mathbf{k}_{+}\mathbf{k}'_{+}}(E;\omega,\vecq) &= L_{\mathbf{k}_{+}\mathbf{k}'_{+}}(E;\omega,\vecq) \frac{\Delta G_\veck(E;\omega,\vecq)}{\Delta\Sigma_\veck(E;\omega,\vecq)}\,,\\[.5em]
\mathcal{G}_{\mathbf{k}_{+}\mathbf{k}'_{+}}(E;\omega,\vecq) &=  \Gamma_{\mathbf{k}_{+}\mathbf{k}'_{+}}(E;\omega,\vecq) \frac{\Delta G_\veck(E;\omega,\vecq)}{\Delta\Sigma_\veck(E;\omega,\vecq)}\,,
\end{align}
for which we formulate the Bethe-Salpeter equation in the low-energy limit. The leading low-energy asymptotics reads
\begin{widetext}
\begin{multline} \label{eq:BS-asymptotic}
\mathcal{G}^{RA}_{\veck_{+}\veck'_{+}}(E;\omega,\vecq) = \mathcal{L}^{RA}_{\veck\veck''} +
\frac 1N \sum_{\veck''}\left\{
\mathcal{L}^{RA}_{\veck\veck''}
\left[1 + \frac{\omega}{\Delta\Sigma_{\mathbf{k}''}}  -\frac{\Delta\epsilon_{\veck''}(\vecq) }{\Delta\Sigma_{\veck''}(E;0,\vecq)}  - \frac{\Delta\epsilon_{\veck''}(\vecq)^{2} }{\Delta\Sigma_{\veck''}^{2}} \right]  
\right. \\ \left.
+ \omega \dot{\mathcal{L}}^{RA}_{\veck\veck''}+ q\hat\vecq\cdot\nabla_{q}\mathcal{L}^{RA}_{\veck_{+}\veck''_{+}}(\vecq)\left(1 - \frac{\Delta\epsilon_{\veck''}(\vecq) }{\Delta\Sigma_{\veck''}}\right)  + \frac {q^{2}}2 (\hat\vecq\cdot\nabla_{q})^{2}\mathcal{L}^{RA}_{\veck_{+}\veck''_{+}}(\vecq) \right\}\mathcal{G}^{RA}_{\veck''\veck'_{+}}(E;\omega,\vecq)\,.
\end{multline}
The functions without the values of their variables are taken at the Fermi energy $E$ for $\omega=i0^{+}$ and $\vecq =0$. Moreover, momentum $\vecq$ at the irreducible vertices is set zero after the $q$ derivatives were performed. After regrouping the terms we obtain
\begin{multline} \label{eq:BS-asymptotic_2}
\mathcal{G}^{RA}_{\veck_{+}\veck'_{+}}(E;\omega,\vecq) = \mathcal{L}^{RA}_{\veck\veck'} +
\frac 1N \sum_{\veck''}\left\{\mathcal{L}^{RA}_{\veck\veck''} + 
\omega\left[\mathcal{L}^{RA}_{\veck\veck''}\frac{1}{\Delta\Sigma_{\mathbf{k}''}}  +\dot{\mathcal{L}}^{RA}_{\veck\veck''} \right]  - q\left[\frac{\hat\vecq\cdot\vecv_{\veck''}}{\Delta\Sigma_{\mathbf{k}''}} - \hat\vecq\cdot\nabla_{q} \right]\mathcal{L}^{RA}_{\veck_{+}\veck''_{+}}(\vecq)
\right. \\ \left.
- q^{2}\left[\frac{\hat\vecq\cdot\vecv_{\veck''}}{\Delta\Sigma_{\mathbf{k}''}}\left(\frac{\hat\vecq\cdot\vecv_{\veck''} - \hat\vecq\cdot\nabla_{q} \Re \Sigma^{R}_{\veck''} }{\Delta\Sigma_{\mathbf{k}''}} + \hat\vecq\cdot\nabla_{q}\right)  - \frac 12 (\hat\vecq\cdot\nabla_{q})^{2} \right]\mathcal{L}^{RA}_{\veck_{+}\veck''_{+}}(\vecq)\right\}\mathcal{G}^{RA}_{\veck''\veck'_{+}}(E;\omega,\vecq)\ .
\end{multline}
We now split the two-particle vertex into two parts, one with even and one with odd symmetry with respect to momentum inversion $q \to -q$.   We gather the odd contributions to form even-symmetry terms that stay in the denominator of both functions. Since we are interested only in the leading non-vanishing order we sum the contributions to the denominator of the full vertex to second order. The corresponding expansion for the even part is
\begin{multline} \label{eq:BS-asymptotic+}
\mathcal{G}^{+}_{\veck_{+}\veck'_{+}}(E;\omega,\vecq) = \mathcal{L}^{RA}_{\veck\veck''} +
\frac 1N \sum_{\veck''}\left\{\mathcal{L}^{RA}_{\veck\veck''} +
\omega\left[\mathcal{L}^{RA}_{\veck\veck''}\frac{1}{\Delta\Sigma_{\mathbf{k}''}}  +\dot{\mathcal{L}}^{RA}_{\veck\veck''} \right]  
\right. \\ \left.  
+ \frac {q^{2}}{N^{2}}\sum_{\veck_{1}\veck_{2}}\mathcal{L}^{RA}_{\veck_{+}\veck_{1+}}(\vecq)\left(\overleftarrow{\nabla}_{q}\cdot\hat\vecq  -\frac{\hat\vecq\cdot\vecv_{\veck_{1}}}{\Delta\Sigma_{\mathbf{k}_{1}}} \right)\left[\widehat{1} - \widehat{\mathcal{L}}^{RA}(\omega) \right]^{-1}_{\veck_{1}\veck_{2}}\mathcal{L}^{RA}_{\veck_{2+}\veck''_{+}}(\vecq)\left(\overleftarrow{\nabla}_{q}\cdot\hat\vecq - \frac{\hat\vecq\cdot\vecv_{\veck''}}{\Delta\Sigma_{\mathbf{k}''}}  \right)
\right. \\ \left.
- q^{2}\mathcal{L}^{RA}_{\veck_{+}\veck''_{+}}(\vecq)\left[\left(\overleftarrow{\nabla}_{q}\cdot\hat\vecq + \frac{\hat\vecq\cdot\vecv_{\veck''} - \hat\vecq\cdot\nabla \Re \Sigma^{R}_{\veck''} }{\Delta\Sigma_{\mathbf{k}''}}\right)\frac{\hat\vecq\cdot\vecv_{\veck''}}{\Delta\Sigma_{\mathbf{k}''}}  - \frac 12 (\overleftarrow{\nabla}_{q}\cdot\hat\vecq)^{2} \right]
\right\}\mathcal{G}^{+}_{\veck''\veck'_{+}}(E;\omega,\vecq)\, ,
\end{multline}
while that for the odd one reads
\begin{multline} \label{eq:BS-asymptotic-}
\mathcal{G}^{-}_{\veck_{+}\veck'_{+}}(E;\omega,\vecq) = \frac {q}{N}\sum_{\veck''} \mathcal{L}^{RA}_{\veck_{+}\veck''_{+}}(\vecq)\left(\overleftarrow{\nabla}_{q}\cdot\hat\vecq - \frac{\hat\vecq\cdot\vecv_{\veck''}}{\Delta\Sigma_{\mathbf{k}''}} \right)\mathcal{G}^{RA}_{\veck''\veck'}(\omega) +\frac {1}{N}\sum_{\veck''}\left\{\mathcal{L}^{RA}_{\veck\veck''} + 
\omega\left[\mathcal{L}^{RA}_{\veck\veck''}\frac{1}{\Delta\Sigma_{\mathbf{k}''}} 
\right.\right. \\ \left.\left.
  +\dot{\mathcal{L}}^{RA}_{\veck\veck''} \right]     + \frac {q^{2}}{N^{2}}\sum_{\veck_{1}\veck_{2}}\mathcal{L}^{RA}_{\veck_{+}\veck_{1+}}(\vecq)\left( \overleftarrow{\nabla}_{q}\cdot\hat\vecq - \frac{\hat\vecq\cdot\vecv_{\veck_{1}}}{\Delta\Sigma_{\mathbf{k}_{1}}}\right)\left[\widehat{1} - \widehat{\mathcal{L}}^{RA}(\omega) \right]^{-1}_{\veck_{1}\veck_{2}}\mathcal{L}^{RA}_{\veck_{2+}\veck''_{+}}(\vecq)\left( \overleftarrow{\nabla}_{q}\cdot\hat\vecq - \frac{\hat\vecq\cdot\vecv_{\veck''}}{\Delta\Sigma_{\mathbf{k}''}}\right)
\right. \\ \left.
- q^{2}\mathcal{L}^{RA}_{\veck_{+}\veck''_{+}}(\vecq)\left[\left( \overleftarrow{\nabla}_{q}\cdot\hat\vecq + \frac{\hat\vecq\cdot\vecv_{\veck''} - \hat\vecq\cdot\nabla \Re \Sigma^{R}_{\veck''} }{\Delta\Sigma_{\mathbf{k}''}}\right)\frac{\hat\vecq\cdot\vecv_{\veck''}}{\Delta\Sigma_{\mathbf{k}''}}  - \frac 12 (\overleftarrow{\nabla}_{q}\cdot\hat\vecq)^{2} \right]
\right\}\mathcal{G}^{-}_{\veck''\veck'_{+}}(E;\omega,\vecq) \,.
\end{multline}
\end{widetext}
The inverse of operator $\widehat{1} - \widehat{\mathcal{L}}^{RA}$ should be taken as the limit $\lim_{\omega \to 0}\bigl[\,\widehat{1} - \widehat{\mathcal{L}}^{RA}(\omega)\bigr]^{-1}$ in order to regularize its low-energy pole. The same holds also for vertex $\mathcal{G}^{RA}(\omega)$. The term containing the inversion $\bigl[\,\widehat{1} - \widehat{\mathcal{L}}^{RA}(\omega) \bigr]^{-1}$ is dominant in the small frequency and small momentum expansion of the irreducible vertex $\mathcal{L}_{\mathbf{k}_{+}\mathbf{k}'_{+}}(E;\omega,\vecq)$ when applied in low spatial dimensions. 

The asymptotic expressions will now be used to determine the low-energy behavior of the electron-hole correlation function and to find exact representations of the the diffusion pole and the diffusion constant.

\subsection{Low-frequency limit of the homogeneous two-particle vertex}

We first set $\vecq =0$ and investigate the low-frequency limit of the full vertex $\Gamma_{\veck\veck'}(\omega)$. The homogeneous two-particle vertex is then determined from a Bethe-Salpeter equation 
\begin{multline}\label{eq:BS-homegeneous}
\Gamma_{\mathbf{k}\mathbf{k}'}(\omega) 
= L_{\mathbf{k}\mathbf{k}'}(\omega) 
\\
+ \frac 1N\sum_{\mathbf{k}''}  \frac{L_{\mathbf{k}\mathbf{k}''}(\omega)\Delta G_{\mathbf{k}''}(\omega)}{\Delta \Sigma_{\mathbf{k}''}(\omega) - \omega}\Gamma_{\mathbf{k}''\mathbf{k}'}(\omega) \,,
 \end{multline}
 where we skipped the superscript $RA$ at the two-particle vertices.
It is straightforward to derive the following low-frequency asymptotics 
\begin{subequations}
\begin{align}
L_{\mathbf{k}\mathbf{k}'}(\omega) &= L_{\mathbf{k}\mathbf{k}'} + i\omega \Im \dot{L}_{\mathbf{k}\mathbf{k}'} \ , \\
\Im\dot{L}_{\mathbf{k}\mathbf{k}'} &= \Im\dot{\Lambda}_{\mathbf{k}\mathbf{k}'} + \frac{1}{2\left\langle \Im G^{2}\right\rangle}\left[\Im G_{\veck}\dot{R}_{\veck'} 
 \right. \nonumber \\ & \left. 
 + \dot{R}_{\veck}\Im G_{\veck'}
 - \left\langle \dot{R} \Im G\right\rangle \frac{\Im G_{\veck}\Im G_{\veck'}}{\left\langle \Im G^{2}\right\rangle}  \right]
 \ , 
 \end{align}
 \begin{align}
\dot{R}_{\veck}&= - \frac 2N\sum_{\veck'} \Im\dot{\Lambda}_{\mathbf{k}\mathbf{k}'}\Im G_{\veck'} 
 \nonumber \\ & \qquad 
 + \frac 1N\sum_{\veck'}\Lambda_{\mathbf{k}\mathbf{k}'}\Re \dot{G}_{\veck'}  - \Re\dot{\Sigma}_{\veck} \ , \\  
\Delta G_{\mathbf{k}}(\omega) &= 2i \Im G_{\mathbf{k}}\left[1 + \omega \frac{\Re \dot{G}_{\mathbf{k}}}{\Im G_{\mathbf{k}}} \right] \  , 
\\
\Delta \Sigma_{\mathbf{k}}(\omega) &= 2i \Im \Sigma_{\mathbf{k}}\left[1 + \omega \frac{\Re \dot{\Sigma}_{\mathbf{k}}}{\Im \Sigma_{\mathbf{k}}} \right] \ , 
\\
\frac{\Delta G_{\mathbf{k}''}(\omega)}{\Delta \Sigma_{\mathbf{k}''}(\omega) - \omega}  &= \frac{\Im G_{\mathbf{k}}}{\Im \Sigma_{\mathbf{k}}}\left[ 1 + \frac{\omega}{2i \Im\Sigma_{\mathbf{k}}} 
\right. \nonumber \\ 
&\left. \times\left(\frac{\Im \Sigma_{\mathbf{k}}}{\Im G_{\mathbf{k}}} \Re \dot{G}_{\mathbf{k}} - \Re \dot{\Sigma}_{\mathbf{k}} + 1\right)\right]\ .
\end{align}\end{subequations}
Using these results we expand Eq.~\eqref{eq:BS-homegeneous} up to the leading small-frequency term
\begin{multline}
\Gamma_{\mathbf{k}\mathbf{k}'}(\omega) = L_{\mathbf{k}\mathbf{k}'} + \frac 1N\sum_{\mathbf{k}''} L_{\mathbf{k}\mathbf{k}''} 
\frac{\Im G_{\mathbf{k}''}}{\Im \Sigma_{\mathbf{k}''}} \Gamma_{\mathbf{k}''\mathbf{k}'}(\omega) 
\\
+ \ \frac{i\omega}{N}\sum_{\mathbf{k}''}\left[ \Im \dot{L}_{\mathbf{k}\mathbf{k}''} -  \frac {L_{\mathbf{k}\mathbf{k}''}}{2\Im \Sigma_{\mathbf{k}''}}\left( \frac{\Im \Sigma_{\mathbf{k}''}}{\Im G_{\mathbf{k}''}} \Re \dot{G}_{\mathbf{k}''} - \Re \dot{\Sigma}_{\mathbf{k}''} + 1\right)\right] 
\\
\times\frac{\Im G_{\mathbf{k}''}}{\Im \Sigma_{\mathbf{k}''}} \Gamma_{\mathbf{k}''\mathbf{k}'}(\omega) \,.
\end{multline}
We skipped the variables if their values were zero and the functions were evaluated at  the Fermi energy for $\omega= i0^{+}$ and $q=0$. 

The operator (matrix) form of the Bethe-Salpeter equation for vertex $\bar{\mathcal{G}}(\omega)$ reads
\begin{equation}\label{eq:G-asymmetric}
\widehat{\mathcal{G}}(\omega) = \widehat{\mathcal{L}} + \widehat{\mathcal{L}} \widehat{\mathcal{G}} + i\omega\left[ \Im \widehat{\dot{\mathcal{L}}}  - \mathcal{L}\frac 1{2\Im \Sigma} \right]\widehat{\mathcal{G}}(\omega) 
\end{equation}
with 
\begin{multline}
 \Im\widehat{\dot{\mathcal{L}}}_{\mathbf{k}\mathbf{k}'}
 \\
  =  \left[\Im \dot{L}_{\mathbf{k}\mathbf{k}'} - \frac {L_{\mathbf{k}\mathbf{k}'}}2 \left( \frac{ \Re \dot{G}_{\mathbf{k}'}}{\Im G_{\mathbf{k}'}} - \frac {\Re \dot{\Sigma}_{\mathbf{k}'} }{\Im \Sigma_{\mathbf{k}'}}\right) \right] \frac{\Im G_{\mathbf{k}'}}{\Im \Sigma_{\mathbf{k}'}}\,.
\end{multline}
We demonstrate the existence of a pole in a specific matrix element of vertex $\widehat{\mathcal{G}}(\omega)$ in the limit $\omega\to0$. We multiply Eq.~\eqref{eq:G-asymmetric} by a left normalized vector $\langle \Im G\rvert =(N\sqrt{\langle \Im G^{2}\rangle})^{-1}\sum_{\veck}\Im G_{\veck}\langle \veck \rvert$ and sum over the right momentum in the two-particle vertices. If we introduce 
$\lvert \mathcal{G}(\omega)\rangle = N^{-2}\sum_{\veck,\veck'}\lvert \veck\rangle \mathcal{G}_{\veck\veck'}(\omega)$ we obtain
\begin{multline}\label{eq:IG-Gomega0}
\left\langle \Im G \rvert \mathcal{G}(\omega) \right\rangle =  \frac{\left\langle \Im G \right\rangle}{\sqrt{\left\langle \Im G^{2}\right\rangle}} +  \left\langle \Im G \rvert \mathcal{G}(\omega) \right\rangle 
\\
 + i \omega \left\langle \Im G \left\lvert\left[\Im \widehat{\dot{\mathcal{L}}} - \mathcal{L}\frac 1{2\Im \Sigma}\right]\right\rvert\mathcal{G}(\omega)\right\rangle \ .
\end{multline}
We assume that $\langle \Im G\rvert$ is the only maximal left eigenvector of operator $\widehat{\mathcal{L}}$. We choose an appropriate orthonormal basis with the maximal eigenvector in the one-particle Hilbert space, that is, we have a decomposition of the unity operator $\widehat{1} = \lvert \Im G\rangle\langle \Im G\rvert  +  \sum_{n} \left\lvert \phi_{n} \right\rangle \left\langle \phi_{n}\right\rvert$. We  evaluate projections of vertex $\widehat{\mathcal{G}}(\omega)$ to the basis vectors $\lvert \phi_{n}\rangle$ and obtain 
\begin{multline}\label{eq:Phin-1}
\left\langle \phi_{n}\rvert \mathcal{G}(\omega) \right\rangle = \left\langle \phi_{n}\rvert \mathcal{L} \right\rangle +  \left\langle \phi_{n} \left\lvert  \widehat{\mathcal{L}} \right\rvert\mathcal{G}(\omega) \right\rangle  
\\
+ i \omega \left\langle \phi_{n} \left\lvert\left[  \Im \widehat{\dot{\mathcal{L}}} - \mathcal{L}\frac 1{2\Im \Sigma}\right]\right\rvert\mathcal{G}(\omega)\right\rangle  
\doteq \left\langle \phi_{n} \left\lvert  \widehat{\mathcal{L}} \right\rvert \Im G\right\rangle
\\
\times  \left\langle \Im G \rvert \mathcal{G}(\omega) \right\rangle + \sum_{m} \left\langle \phi_{n} \left\lvert  \widehat{\mathcal{L}} \right\rvert \phi_{m}\right\rangle \left\langle \phi_{m}\rvert \mathcal{G}(\omega) \right\rangle
\ .
\end{multline}
We  could neglect the first term on the right-hand side and, since we assumed that the operator of the frequency derivative $\Im \widehat{\dot{\mathcal{L}}}$ is bounded, also  the frequency-dependent term in the second line. They are nonsingular. The absolute term may be divergent via a projection onto the eigenvector $\lvert \Im G\rangle$. Its divergent part reads: 
\begin{multline}
\left\langle \phi_{n}\rvert \mathcal{G}(\omega) \right\rangle 
= \sum_{m}\left\langle \phi_{n}\left\lvert \left(\widehat{1}  - \widehat{P}\widehat{\mathcal{L}}\right)^{-1}\right\rvert \phi_{m}\right\rangle 
\\
\times \left\langle \phi_{m}\left\lvert \widehat{\mathcal{L}} 
\right\rvert\Im G \right\rangle \left\langle \Im G \rvert \mathcal{G}(\omega) \right\rangle 
\end{multline}
where we denoted projectors $\widehat{P}= \sum_{n} \left\lvert \phi_{n}\right\rangle\left\langle \phi_{n}\right\rvert = \widehat{1} - \widehat{Q} = \widehat{1} - \lvert \Im G \rangle\langle \Im G\rvert$ and used an eigenvalue equation $\widehat{Q}\widehat{\mathcal{L}} =\widehat{Q} $ together with the orthogonality relations $\langle \phi_{n}\lvert \Im G\rangle = 0$.  

After summing over the intermediate states and using Eq.~\eqref{eq:Phin-1} we obtain
\begin{multline}\label{eq:Phi-Gomega}
\frac{\left\langle \phi_{n} \rvert \mathcal{G}(\omega) \right\rangle}{ \left\langle \Im G \rvert \mathcal{G}(\omega) \right\rangle} 
= \left\langle \phi_{n}\left\lvert\left( \widehat{1} - \widehat{P}\widehat{\mathcal{L}}  \right)^{-1}\widehat{P}\widehat{\mathcal{L}}\right\rvert\Im G \right\rangle
\\
 =\left\langle \phi_{n}\left\lvert\left( \widehat{1} - \widehat{\mathcal{L}} + \widehat{Q}\right)^{-1}\left( \widehat{\mathcal{L}} - \widehat{Q}\right)\right\rvert\Im G \right\rangle  \,.
\end{multline}
The above equation can be rewritten in a vector form
\begin{equation}\label{eq:P-CalG}
\widehat{P}\left\lvert \mathcal{G}(\omega)\right\rangle = \left( \widehat{1} - \widehat{P}\widehat{\mathcal{L}}  \right)^{-1}\widehat{P}\widehat{\mathcal{L}}\widehat{Q}\left\lvert\Im G \right\rangle \ .
\end{equation}
Further on we use decoupling $\lvert \mathcal{G}(\omega)\rangle = [\widehat{P} + \widehat{Q}]\lvert \mathcal{G}(\omega)\rangle$, Eq.~\eqref{eq:P-CalG}, and $\widehat{Q}\lvert \mathcal{G}(\omega)\rangle = \alpha(\omega) \lvert \Im G\rangle$ to obtain
\begin{widetext}
\begin{equation}
\left\lvert \mathcal{G}(\omega)\right\rangle = \left( \widehat{1} - \widehat{\mathcal{L}} + \widehat{Q}\right)^{-1} \widehat{Q}\left\lvert \mathcal{G}(\omega)\right\rangle
= \alpha(\omega)  \left( \widehat{1} - \widehat{\mathcal{L}} + \widehat{Q}\right)^{-1}\left\lvert \Im G\right\rangle\ . 
\end{equation}
We use  Eq.~\eqref{eq:IG-Gomega0} to determine the unknown function $\alpha(\omega)$ 
\begin{align}
 \frac{\left\langle \Im G \right\rangle}{\sqrt{\left\langle \Im G^{2}\right\rangle}} &= -i\omega\left[\left\langle \Im G \left\lvert\left[ \Im \widehat{\dot{\mathcal{L}}} - \mathcal{L}\frac 1{2\Im \Sigma} \right]\right\rvert\Im G \right\rangle \left\langle \Im G\lvert \mathcal{G}(\omega)\right\rangle + \sum_{n}\left\langle \Im G \left\lvert\left[ \Im \widehat{\dot{\mathcal{L}}} - \mathcal{L}\frac 1{2\Im \Sigma} \right]\right\rvert\phi_{n} \right\rangle \left\langle \phi_{n}\lvert \mathcal{G}(\omega)\right\rangle\right] \nonumber\\
 & = -i\omega\left\langle \Im G \left\lvert\left[ \Im \widehat{\dot{\mathcal{L}}} - \mathcal{L}\frac 1{2\Im \Sigma}\right] \left[\widehat{1} + \left(\widehat{1} - \widehat{P} \widehat{\mathcal{L}} \right)^{-1}\widehat{P} \widehat{\mathcal{L}} \right] \right\rvert\Im G \right\rangle \alpha(\omega)\,.
\end{align}
The final expression for the singular part of vector $\left\lvert \mathcal{G}(\omega)\right\rangle$ then reads
\begin{equation}\label{eq:CalG-reduced}  
\left\lvert \mathcal{G}(\omega)\right\rangle = \frac{\left\langle \Im G \right\rangle}{- i \omega  \sqrt{\left\langle \Im G^{2}\right\rangle}  \left\langle \Im G \left\lvert \left[ \Im \widehat{\dot{\mathcal{L}}} - \mathcal{L}\displaystyle{\frac 1{2\Im \Sigma}} \right] \left(\widehat{1} -  \widehat{\mathcal{L}} + \widehat{Q}\right)^{-1} \right\rvert\Im G \right\rangle}\ \left(\widehat{1} -  \widehat{\mathcal{L}} + \widehat{Q}\right)^{-1} \left\lvert\Im G \right\rangle \ .
\end{equation}

Next, we rewrite the denominator in Eq.~\eqref{eq:CalG-reduced} in terms of the original two-particle vertex $\Lambda$. The expression in the square brackets can be written as
\begin{multline}\label{eq:IG-Gomega-final}
\Im\dot{\mathcal{L}}_{\veck\veck'} - \mathcal{L}_{\veck\veck'}\frac 1{2\Im\Sigma_{\veck'}} = \left\{\Im\dot{\Lambda}_{\veck\veck'} - \frac 12\Lambda_{\veck\veck'} \left(\frac{\Re \dot{G}_{\veck'}}{\Im G_{\veck'}} - \frac{\Re \dot{\Sigma}_{\veck'}}{\Im \Sigma_{\veck'}}  + \frac{1}{\Im \Sigma_{\veck'}}\right) + \frac{1}{2\left\langle \Im G_{\veck''}^{2}\right\rangle_{\veck''}}
\right. \\ \left.
 \times\left[\Im G_{\veck}\left(\left\langle \Lambda_{\veck'\veck''}\Re \dot{G}_{\veck''} - 2\dot{\Lambda}_{\veck'\veck''}\Im G_{\veck''}\right\rangle_{\veck''} - \Re\dot{\Sigma}_{\veck'}\right)  + \left(\left\langle \Lambda_{\veck\veck''}\Re \dot{G}_{\veck''} - 2\dot{\Lambda}_{\veck\veck''}\Im G_{\veck''}\right\rangle_{\veck''}   - \Re\dot{\Sigma}_{\veck}  \right)\Im G_{\veck'}
 \right.\right. \\ \left.\left.
 + \frac{\Im G_{\veck}\Im G_{\veck'}}{\left\langle \Im G_{\veck''}^{2}\right\rangle_{\veck''}}\left(2\left\langle \Im G_{\veck''} \dot{\Lambda}_{\veck''\veck'''}\Im G_{\veck'''}\right\rangle_{\veck''\veck'''} - \left\langle \Im G_{\veck''} \Lambda_{\veck''\veck'''}\Re \dot{G}_{\veck'''}\right\rangle_{\veck''\veck'''} + \left\langle \Im G_{\veck''} \Re\dot{\Sigma}_{\veck''}\right\rangle_{\veck''} \right) \right]\right\} \frac{\Im G_{\veck'}}{\Im \Sigma_{\veck'}}\, ,
\end{multline}
\end{widetext}
where we employed a compact notation for momentum sums
$N^{-1}\sum_{\veck} f_{\veck} =\langle f_{\veck}\rangle_{\veck}$.
Expanding the inverse operator $\bigl(\,\widehat{1} -  \widehat{\mathcal{L}} + \widehat{Q}\bigr)^{-1}$ in powers of $\widehat{\mathcal{L}} - \widehat{Q}$ and using the properties $\widehat{Q}\rvert\Im G \rangle = \rvert\Im G \rangle $, $\widehat{Q}\widehat{\mathcal{L}} = \widehat{Q}$, and the static Ward identity $ N^{-1}\sum_{\veck'}\mathcal{L}_{\veck\veck'}\Im\Sigma_{\veck'} = \Im\Sigma_{\veck}$, we can explicitly evaluate
%
$\bigl[\, \widehat{1} - \widehat{\mathcal{L}} + \widehat{Q}\bigr]^{-1}\left\lvert\Im G\right\rangle = \lim_{n\to\infty} \mathcal{L}^{n}\left\lvert \Im G\right\rangle
= \left\lvert \Im\Sigma\right\rangle \left\langle \Im\Sigma\lvert\Im G\right\rangle
$. We introduced a unit vector $\left\lvert \Im\Sigma\right\rangle = (N\sqrt{\left\langle \Im \Sigma^{2}\right\rangle})^{-1}\sum_{\veck}\lvert \veck \rangle \Im\Sigma_{\veck}$  in the last step. After a few additional manipulations we come to 
\begin{multline}\label{eq:L-derivative-final}
\left\langle \Im G\left\lvert\left[\Im\dot{\mathcal{L}} - \mathcal{L}\frac 1{2\Im\Sigma} \right]\left(\widehat{1} -  \widehat{\mathcal{L}} + \widehat{Q}\right)^{-1} \right\rvert \Im G \right\rangle 
\\
= \left\langle \Im G\left\lvert\left[\Im\dot{\mathcal{L}} - \mathcal{L}\frac 1{2\Im\Sigma} \right] \right\rvert \Im \Sigma \right\rangle\left\langle \Im \Sigma \lvert \Im G \right\rangle 
\\
= - \frac{\left\langle \Im\Sigma_{\veck}\Im G_{\veck}\right\rangle_{\veck}\left\langle \Im G_{\veck}\right\rangle_{\veck}}{2\left\langle \Im G_{\veck}^{2}\right\rangle_{\veck}\left\langle \Im \Sigma_{\veck}^{2}\right\rangle_{\veck}} 
\end{multline}
and the final expression for the singular part of vertex $\widehat{\mathcal{G}}(\omega)$  that has a rather simple form
\begin{equation}\label{eq:Gcalomega}
\widetilde{\mathcal{G}}^{RA}_{\veck}(\omega) = \frac{2 \Im\Sigma^{R}_{\veck}}{i\omega}\, .
\end{equation}
 Notice that the low-frequency singularity of the homogeneous two-particle vertex $\Gamma_{\mathbf{k}\mathbf{k}'}(\omega)$ is fully controlled by only one-particle functions.  Equation~\eqref{eq:Gcalomega} is an alternative version of the Velick\'y Ward identity.\cite{Velicky:1969aa,Janis03}  The genuine two-particle behavior of the two-particle vertex becomes evident only in the $q$-dependent terms.

\section{Diffusion pole and static diffusion constant}

We use the singularity in the homogeneous vertex $\widehat{\mathcal{G}}(\omega)$ to determine the exact form of the diffusion pole and the diffusion constant controlling the low-energy asymptotics of the electron-hole correlation function.

\subsection{Derivation of the diffusion pole}

The uncorrelated part of the electron-hole correlation function $ \Phi(E;\omega,\vecq)$
is regular and hence the pole  can emerge only in the vertex contribution
\begin{multline}\label{eq:Phi-vertex-part}
\frac1{N^2}\sum_{\veck\veck'}
\GGq{} \Gamma^{RA}_{{\bf k}_{+}{\bf k}'_{+}}(E,\omega;{\bf q})
\\
\times  \GGq{'}
 \end{multline}
that will be found by summing the Bethe--Salpeter equation~\eqref{eq:BS-fundamental}
over the fermionic momenta $\veck$ and $\veck'$. 

We first multiply Eq.~\eqref{eq:BS-fundamental} with $\Delta
G_{\veck'}(E;\omega,\mathbf q)/\Delta\Sigma_{\veck'}(E;\omega,\mathbf q)$ and
sum over $\veck'$, which yields
\begin{multline}\label{eq:BS-sum1}
\mathcal{G}^{RA}_{\veck}(E;\omega,\vecq)=
\biggl\langle L^{RA}_{\veck_{+}\veck''_{+}}(E;\omega,\vecq) \biggl[
  \frac{\Delta G_{\veck''}(E;\omega,\vecq)}{%
    \Delta\Sigma_{\veck''}(E;\omega,\vecq)}
    \\
+ \GGq{''}\, \mathcal{G}^{RA}_{\veck''}(E;\omega,\vecq)\biggr]
\biggr\rangle_{\veck''}\,.
\end{multline}
We introduced a reduced two-particle function
$\mathcal{G}^{RA}_{\veck}(E;\omega,\vecq)=
\langle
\mathcal{G}^{RA}_{\veck_{+}\veck'}(E;\omega,\vecq)\rangle_{\mathbf{k'}}$.
From now on, we will shorten $\Delta G_{\veck}(E;0,\mathbf 0)$ to
$\Delta G_{\veck}$ and similarly for $\Delta\Sigma_\veck$ unless their
full form is useful to indicate the lack of dependence on $\omega$ or
$\vecq$.

In the next step we multiply Eq.~\eqref{eq:BS-sum1} with
$\Delta G_{\veck}(E;\omega,\vecq)$,
sum over $\veck$, and use the dynamical Ward identity
$\Delta\Sigma_{\mathbf{k''}}(E;\omega,\vecq)=\bigl\langle
L^{RA}_{\veck\veck''}(E;\omega,\vecq)
\Delta G_{\veck}(E;\omega,\vecq)\bigr\rangle_{\veck}$
to eliminate the irreducible vertex $L^{RA}$.  We arrive at an
integral equation for $\mathcal{G}^{RA}_{\veck}(E;\omega,\vecq)$,
\begin{multline}
\Bigl\langle \Bigl[
\Delta G_{\veck}(E;\omega,\vecq)
- \Delta\Sigma_{\veck}(E;\omega,\vecq)
    \GGq{} \Bigr]
    \\
    \times 
    \mathcal{G}^{RA}_{\veck}(E;\omega,\vecq)
\Bigr\rangle_{\veck}=
\biggl\langle \Delta G_{\veck}\,
  \frac{\Delta\Sigma_\veck(E;\omega,\vecq)}{%
  \Delta\Sigma_\veck}
\biggr\rangle_{\veck}\,.
\label{eq:summedBSbeforeQexp}
\end{multline}

The product of two Green functions on the left-hand side is further
rewritten in terms of differences, Eq.~\eqref{eq:GG-DeltaG}, which yields
\begin{multline}\label{eq:Gcal-reduced}
\Bigl\langle
\frac{\left[-\omega+\Delta\epsilon_{\veck}(\vecq)\right]\Delta G_{\veck}(E;\omega,\vecq)}{\Delta\Sigma_{\veck}(E;\omega,\vecq) - \omega
 + \Delta\epsilon_{\veck}(\vecq)}\,
  \mathcal{G}^{RA}_{\veck}(E;\omega,\vecq)
\Bigr\rangle_{\veck}
\\
=  \biggl\langle \Delta G_{\veck}\,
  \frac{\Delta\Sigma_\veck(E;\omega,\vecq)}{%
  \Delta\Sigma_\veck}
\biggr\rangle_{\veck}\,.
\end{multline}
Since we are interested in the behavior of
$\mathcal{G}^{RA}_{\veck}(E;\omega,\vecq)$ only at small $q$ and
small $\omega$, it is sufficient to expand the factor standing at
$\mathcal{G}^{RA}_{\veck}(E;\omega,\vecq)$ in powers of $q$ and
$\omega$, and keep only terms up to $q^2$ and
$\omega$. Higher-order terms, including $\omega q$, will be
neglected. Because we expect
$\mathcal{G}^{RA}_{\veck}(E;\omega,\vecq)$ to be singular at small $q$
and $\omega$, and because the expression on the right-hand side is
regular, we can set $\omega=0$ and $\vecq=\mathbf 0$ there right
away. Furthermore, taking 
into account that $\Delta\epsilon_{\veck}(\vecq)$ is
proportional to~$q$, we can immediately write
\begin{multline}
\label{eq:summedBS-omega-done}
-\omega\,\Bigl\langle
\frac{\Delta G_{\veck}(E;0,\mathbf 0)}{%
\Delta\Sigma_{\veck}(E;0,\mathbf 0)}\,
  \mathcal{G}^{RA}_{\veck}(E;\omega,\vecq)
\Bigr\rangle_{\veck}\\
+\Bigl\langle\Delta\epsilon_{\veck}(\vecq)\,
\frac{\Delta G_{\veck}(E;0,\vecq)}{%
\Delta\Sigma_{\veck}(E;0,\vecq)
 + \Delta\epsilon_{\veck}(\vecq)}\,
  \mathcal{G}^{RA}_{\veck}(E;\omega,\vecq)
\Bigr\rangle_{\veck}
\\
= \bigl\langle \Delta G_{\veck}
\bigr\rangle_{\veck}\,.
\end{multline}
We expand the multiplication factor  at
$\mathcal{G}^{RA}_{\veck}(E;\omega,\vecq)$ in Eq.~\eqref{eq:summedBS-omega-done}  in powers of $q=|\vecq|$. Since this function
is already proportional to $\Delta\epsilon_{\veck}(\vecq)$, we
only need to expand $\Delta\epsilon_{\veck}(\vecq)$, $\Delta
G_{\veck}(E;0,\vecq)$ and $\Delta\Sigma_{\veck}(E;0,\vecq)$
up to terms linear in~$q$. 


The possibility to express everything in terms of real and imaginary
parts of the retarded Green function and retarded self-energy comes
from the fact that the argument $z$ in
Eq.~\eqref{eq:qexpansion-generic} is always $z=E\pm\rmi0$. Inserting
Eqs.~\eqref{eq:qexpansion-components} into the second term of
Eq.~\eqref{eq:summedBS-omega-done} we arrive at
\begin{multline}
\frac{\Delta\epsilon_{\veck}(\vecq)\,
\Delta G_{\veck}(E;0,\vecq)}{%
\Delta\Sigma_{\veck}(E;0,\vecq)
 + \Delta\epsilon_{\veck}(\vecq)}=
q(\hat\vecq\cdot\vecv_\veck)|G_\veck^R|^2 +\\
\underbrace{\rmi q^2 (\hat\vecq\cdot\vecv_\veck)|G_\veck^R|^2\bigl[
  (\hat\vecq\cdot\vecv_\veck)\Im G_\veck^R
  +\Im\bigl(G_\veck^R(\hat\vecq\cdot\nabla_\veck)\Sigma_\veck^R
  \bigr)\bigr]}_{\displaystyle -\rmi q^2\mathcal D_\veck}\,,
\end{multline}
where we repeatedly used Eq.~\eqref{eq:GG-DeltaG} at $\omega=0$ and
$\vecq=\mathbf 0$, that is,
$\Delta G_\veck/\Delta\Sigma_\veck=|G_\veck^R|^2$. Coming back to
Eq.~\eqref{eq:summedBS-omega-done} we have
\begin{multline}
\label{eq:summedBS-with-q-linear}
-\omega\,\widetilde\Phi(E;\omega,\mathbf q)
+q\,\bigl\langle (\hat\vecq\cdot\vecv_\veck)|G_\veck^R|^2
  \mathcal{G}^{RA}_{\veck}(E;\omega,\vecq)
  \bigr\rangle_{\veck}
  \\
-\rmi q^2 \bigl\langle
  \mathcal D_\veck \mathcal{G}^{RA}_{\veck}(E;\omega,\vecq)
  \bigr\rangle_{\veck}
=\bigl\langle \Delta G_{\veck}\bigr\rangle_{\veck}\,,
\end{multline}
where we introduced a reduced electron-hole correlation function 
\begin{multline}\label{eq:PhiTilde-def}
\widetilde\Phi(E;\omega,\mathbf q)
=\bigl\langle
  |G_\veck^R|^2\mathcal{G}^{RA}_{\veck}(E;\omega,\vecq)
  \bigr\rangle_{\veck}
=\\\bigl\langle
  |G_\veck^R|^2
  \Gamma^{RA}_{\veck\veck'}(E;\omega,\vecq)|G_{\veck'}^R|^2
  \bigr\rangle_{\veck\veck'}\,.
\end{multline}
The right-hand side of Eq.~\eqref{eq:summedBS-with-q-linear} can be evaluated with
$\bigl\langle \Delta G_{\veck}\bigr\rangle_{\veck}=
2\rmi\bigl\langle\Im G_{\veck}^R\bigr\rangle_{\veck}=
-2\pi\rmi\, n_{\rm F}$. For $\vecq=\mathbf 0$ we straightforwardly find
\begin{equation}\label{eq:Phi-omega}
\widetilde\Phi(\omega)=\bigl\langle
  |G_\veck^R|^2\mathcal{G}^{RA}_{\veck}(\omega)
  \bigr\rangle_{\veck}=
\frac{2\pi n_{\rm F}}{-\rmi\omega}\,.
\end{equation}

For non-vanishing momentum $\vecq$, we can rewrite Eq.~\eqref{eq:summedBS-with-q-linear} 
to
\begin{widetext}
\begin{equation}
\label{eq:finalEq-Phi}
\biggl[-\rmi\omega
+\frac{%
  \rmi q\,\bigl\langle (\hat\vecq\cdot\vecv_\veck)|G_\veck^R|^2
  \mathcal{G}^{-}_{\veck_{+}\veck'_{1}}(E;\omega,\vecq)
  \bigr\rangle_{\veck\veck'}
  +q^2 \bigl\langle
 \mathcal D_\veck\mathcal{G}^{+}_{\veck_{+}\veck'_{+}}(E;\omega,\vecq)
 \bigr\rangle_{\veck\veck'}}{%
 \bigl\langle
  |G_\veck^R|^2\mathcal{G}^{+}_{\veck_{+}\veck'_{+}}(E;\omega,\vecq)
  \bigr\rangle_{\veck\veck'}}
\biggr]\widetilde\Phi(E;\omega,\mathbf q)
=2\pi n_{\rm F}\,
\end{equation}
\end{widetext}
where we used Eqs.~\eqref{eq:PhiTilde-def} and~\eqref{eq:Phi-omega} to represent $\widetilde\Phi(E;\omega,\mathbf q)$ in the denominator on the left-hand side. It does not depend on momentum $q$ in the leading order. We now use the solution for vertices $\mathcal{G}^{\pm}$ from the preceding section to obtain an explicit representation for the low-energy limit of the electron-hole correlation function. Using Eq.~\eqref{eq:BS-asymptotic-}, the odd term in Eq.~\eqref{eq:finalEq-Phi} reads
\begin{multline}
q\,\bigl\langle (\hat\vecq\cdot\vecv_\veck)|G_\veck^R|^2
  \mathcal{G}^{-}_{\veck_{+}\veck'_{+}}(E;\omega,\vecq)
  \bigr\rangle_{\veck\veck'} 
  \\
  = q^{2} \left\langle( \hat\vecq\cdot\vecv_\veck)|G_\veck^R|^2
  \left[\widehat{1} - \widehat{\mathcal{L}} \right]^{-1}_{\veck\veck'} \mathcal{L}^{RA}_{\veck'_{+}\veck''_{+}}(\vecq)\left(\overleftarrow{\nabla}_{q}\cdot\hat\vecq 
  \right.\right. \\ \left.\left.
  - \frac{\hat\vecq\cdot\vecv_{\veck''}}{\Delta\Sigma_{\mathbf{k}''}}\right)\mathcal{G}^{RA}_{\veck''}(\omega)\right\rangle_{\veck\veck'\veck''} 
\end{multline}
which is also proportional to $q^{2}$ and contributes to the diffusion constant. 

We further extract the gradient of the one-electron functions from vertex $\mathcal{L}^{RA}_{\veck\veck'}(\vecq)$. We have
\begin{multline}
i\nabla_{q}\mathcal{L}_{\veck_{+}\veck'_{+}}(q) 
= i\nabla_{q}L_{\veck_{+}\veck'_{+}}(\vecq)\left\lvert G^{R}_{\veck'}\right\rvert^{2}
\\
 + \frac {L_{\veck\veck'}}{2\Im\Sigma^{R}_{\veck'}} 
\left[\left((\Re G^{R}_{\veck'})^{2} - (\Im G^{R}_{\veck'})^{2}\right)\mathbf{v}_{\veck'} 
\right. \\ \left.
- 2 \Im G^{R}_{\veck'}\Im\left( G^{R}_{\veck'}\nabla\Sigma^{R}_{\veck'}\right)\right]\ .
\end{multline} 
Since vertex $L_{\veck\veck'}(\vecq)$ depends only on even powers of momentum $q$, $\nabla_{q}L_{\veck\veck'}(\vecq)=0$.  
With this representation we obtain the singular part of the low-energy limit of the electron-hole correlation function in form of the canonical diffusion pole 
\begin{equation}
\label{eq:Phi-pole}
\widetilde\Phi(E;\omega,\vecq)=
\frac{2\pi n_{\rm F}}{-\rmi\omega+D(\omega)q^2}\ 
\end{equation}
with the diffusion constant 
\begin{multline}\label{eq:DC-dynamical}
D(\omega) = \frac{i\omega}{2\pi n_{F}}\left\langle ( \hat\vecq\cdot\vecv_\veck)|G_\veck^R|^{2}\left[\widehat{1} - \widehat{\mathcal{L}} \right]^{-1}_{\veck\veck'} \left[\Im G_{\veck'}^R\hat\vecq\cdot\vecv_{\veck'} 
\right.\right. \\ \left.\left.
  +\Im\left(G_{\veck'}^R\hat\vecq\cdot\nabla_{\veck'}\Sigma_{\veck'}^R
  \right)\right]  \mathcal{G}^{RA}_{\veck'\veck''}(\omega)\right\rangle_{\veck\veck'\veck''} \ . 
\end{multline} 
Equations~\eqref{eq:Phi-pole} and~\eqref{eq:DC-dynamical} determine the exact canonical form of the diffusion pole in the electron-hole (density-density) correlation function. 

\subsection{Static diffusion constant}

An explicit representation for the static diffusion constant $D=D(0)$, that is, the explicit limit $\omega\to 0$ in Eq.~\eqref{eq:DC-dynamical} is obtained from Eq.~\eqref{eq:DC-dynamical} when we use the singular part of vertex $\mathcal{G}^{RA}_{\veck}(\omega)$ from Eq.~\eqref{eq:Gcalomega}. The static diffusion constant is then expressed via a Kubo-like formula with the full two-particle vertex
\begin{multline}\label{eq:DC-static}
\pi n_{F}D =  \left\langle ( \hat\vecq\cdot\vecv_\veck)|G_\veck|^{2}\left[N\delta_{\veck,\veck'} + \Gamma_{\veck\veck'}\lvert G_{\veck'}\rvert^{2}\right]
\right. \\ \left.
\times \left[\Im G_{\veck'}\hat\vecq\cdot\vecv_{\veck'}  + \Im\left(G_{\veck'}\hat\vecq\cdot\nabla_{\veck'}\Sigma_{\veck'}
  \right)\right] \Im \Sigma_{\veck'}\right\rangle_{\veck\veck'}\, ,
\end{multline}   
where we used $[\widehat{1} - \widehat{\mathcal{L}}]^{-1}_{\veck\veck'} =  N\delta_{\veck,\veck'} + \Gamma_{\veck\veck'}\lvert G_{\veck'}\rvert^{2}$. This exact expression is the starting point for the derivation of consistent approximations for the diffusion constant needed  to reach quantitative results. As a first step we choose the local mean-field approximation for which $\Sigma_{\veck}= \Sigma$ and $\Lambda_{\veck\veck'}= \Im \Sigma/\langle\Im G_{\veck''}\rangle_{\veck''}\equiv\lambda$. Further on, $\mathcal{L}_{\veck\veck'}= \Im G_{\veck'}/\langle \Im G_{\veck''}\rangle_{\veck''}$.
Expression~\eqref{eq:DC-static} for the static diffusion constant then reduces to the CPA result
\begin{equation}\label{eq:DC-CPA}
\pi n_{F}D = \left\langle (\hat\vecq\cdot\vecv_\veck)^{2}\Im G_{\veck}^{2}\right\rangle_{\veck} \ .
\end{equation}
This formula holds for any approximation with a local vertex $\Lambda$. 
 
Approximations with the local irreducible vertex do not improve upon the Drude or CPA diffusion constant. Only a non-local irreducible vertex can make a difference. Since it is complicated to evaluate the full momentum dependence of the self-energy resulting from momentum dependence of the vertex function, we consider only the momentum dependent self-energy as a correction to the local term. We denote
\begin{subequations}\begin{align}
\Sigma^{R}_{\mathbf{k}}(E) & = \Sigma^{R} (E) + \delta\Sigma^{R}_{\mathbf{k}}(E) \ , \\
\Lambda^{RA}_{\mathbf{k}\mathbf{k}'}(E;\omega,\vecq) & = \lambda^{RA}(E;\omega)  + \delta\Lambda^{RA}_{\mathbf{k}\mathbf{k}'}(E;\omega,\mathbf{q}) \ .
\end{align}\end{subequations}
and take the non-local correction from the maximally crossed diagrams, that is
\begin{multline}\label{eq:delta-Lambda}
\delta\Lambda^{RA}_{\mathbf{k}_{+}\mathbf{k}'_{+}} (E;\omega,\vecq) 
\equiv \delta\Lambda(E;\omega,\veck + \veck') 
\\
= \frac{\lambda^{RA}(E;\omega)^{2}\overline{\chi}_{+}(E;\omega, \mathbf{k} + \mathbf{k}')}{1 - \lambda^{RA}(E;\omega)\chi_{+}(E;\omega, \mathbf{k} + \mathbf{k}' )}\ .
\end{multline}
We used the reduced electron-electron bubble
\begin{multline}
\overline{\chi}^{RA}_{+}(E; \omega,\mathbf{q}) = \chi^{RA}_{+}(E; \omega,\mathbf{q}) 
\\
-  G^{R}(E_{+}) G^{A}(E_{-})
= \frac 1N \sum_{\mathbf{k}} G^{R}_{\mathbf{k}_{+}}(E_{+} ) G^{A}_{ - \mathbf{k}_{-}}(E_{-}) 
\\
-  G^{R}(E_{+}) G^{A}(E_{-}) \ .
\end{multline}

The non-local correction to the CPA irreducible vertex leads to a correction to the CPA the self-energy,
\begin{subequations}\begin{align}\label{eq:KK-delta-imaginary}
\Im \delta\Sigma_{\mathbf{k}}(E) & = \frac 1N\sum_{\mathbf{k}'}\delta\Lambda^{RA}_{\mathbf{k}\mathbf{k}'}(E) \Im G_{\mathbf{k}'}(E) \ ,\\ 
\Re \delta\Sigma_{\mathbf{k}}(E) & = P\int_{-\infty}^{\infty} \frac{d\omega}{\pi} \frac{\Im\delta\Sigma_{\mathbf{k}}(\omega)}{\omega - E} \ . \label{eq:KK-delta-real}
\end{align}\end{subequations}
Generally, this correction to the mean field is justified for weak disorder where the momentum dependence of the self-energy is expected to be small. The one-electron propagators contain only the local, unperturbed self-energy.

The non-local irreducible vertex $\delta\Lambda^{RA}_{\veck\veck'}$ depends only on the sum $\veck + \veck'$ and contains the Cooper pole at $\veck + \veck'= 0$. The contribution from this pole to the diffusion constant will be dominant in low dimensions. We use this dominance and also neglect the momentum dependence of the self-energy (the correction $\delta\Sigma_{\veck}$). To determine the diffusion constant, we have to evaluate the following expression
\begin{widetext}
\begin{multline}\label{eq:delta-Lambda-expansion}
v_{\veck}\left[\widehat{1} - \Lambda\left\lvert G\right\rvert^{2} \right]^{-1}_{\veck\veck'}v_{\veck'} = v_{\veck}^{2} N\delta_{\veck,\veck'}  + \sum_{n=1}^{\infty}\frac 1{N^{n-1}}\sum_{\veck_{1}}\sum_{\veck_{2}}\ldots \sum_{\veck_{n}} v_{\veck}\Lambda(\veck + \veck_{1})\lvert G_{\veck_{1}}\rvert^{2}\Lambda(\veck_{1} + \veck_{2})\lvert G_{\veck_{2}}\rvert^{2} \ldots \delta_{\veck_{n},\veck'}v_{\veck'}
\\
= v_{\veck}^{2} N\delta_{\veck,\veck'} +  \sum_{n=1}^{\infty}\frac 1{N^{n-1}}\sum_{\vecq_{1}}\sum_{\vecq_{2}}\ldots \sum_{\vecq_{n}} v_{\veck}\Lambda(\vecq_{1})\lvert G_{\vecq_{1} - \veck}\rvert^{2}\Lambda(\vecq_{2})\lvert G_{\vecq_{2} - \vecq_{1} + \veck}\rvert^{2} \ldots \Lambda(\vecq_{n})\lvert G_{\veck_{n}}\rvert^{2}\delta_{\veck_{n}, \veck'}v_{\veck'}
\end{multline}
\end{widetext}
where $\veck_{n}= \sum_{i=1}^{n}(-1)^{n-i}\vecq_{i} + (-1)^{n}\veck$. Since the vertex correction $\delta\Lambda(\vecq)$ contains the Cooper pole, the dominant contribution to  the integrals in Eq.~\eqref{eq:delta-Lambda-expansion} comes from the values of non-singular functions at $q_{i} =0$. The multiple sums (integrals in the thermodynamic limit) then factorize and we obtain a renormalized diffusion constant in a closed form  
\begin{equation}\label{eq:MFDC-correction}
\pi n_{F} D 
= \left\langle \frac{(\hat\vecq\cdot\vecv_\veck)^{2}\Im G_{\veck}^{2}}{1 + \lvert G_{\veck}\rvert^{2}\left\langle \delta \Lambda^{RA}(q)\right\rangle_{\vecq}}\right\rangle_{\veck} \,,
\end{equation}
where 
\begin{multline}
\left\langle \delta \Lambda^{RA}(q)\right\rangle_{\vecq}
\\
 = \frac 1{\chi^{RA}_{+}(0)}\left[\left\langle \displaystyle{\frac{\chi^{RA}_{+}(0) - \lvert G\rvert^{2}}{\chi^{RA}_{+}(0) - \chi^{RA}_{+}(\vecq)}}\right\rangle_{\vecq} - 1\right]\,.
\end{multline}
The mean-field self-energy $\Sigma^{R}(E)$ is determined from the Soven equation
\begin{equation}\label{eq:SE-CPA}
  1=\left\langle\frac
1{1+\left[\Sigma(E)-V_i\right]G(E)}\right\rangle_{av}
\end{equation}
and the corresponding irreducible vertex is
\begin{multline}\label{eq:2IP-vertex}
  \lambda^{RA}(E) = \frac{\Im\Sigma(E)}{\Im G(E)} = \frac 1{\left\langle \lvert  G_{\veck}(E)\rvert^{2}\right\rangle_{\veck}} =\frac 1{\lvert G(E)\rvert^{2}}
  \\
\times  \left[1 \phantom{\frac 12} - \left\langle \left\lvert\frac 1{1+\left(\Sigma(E) - V_i\right)G(E)}\right\rvert^{2}\right\rangle^{-1}_{av} \right] \,.
\end{multline}

\begin{figure*}
\includegraphics[width=8.5cm]{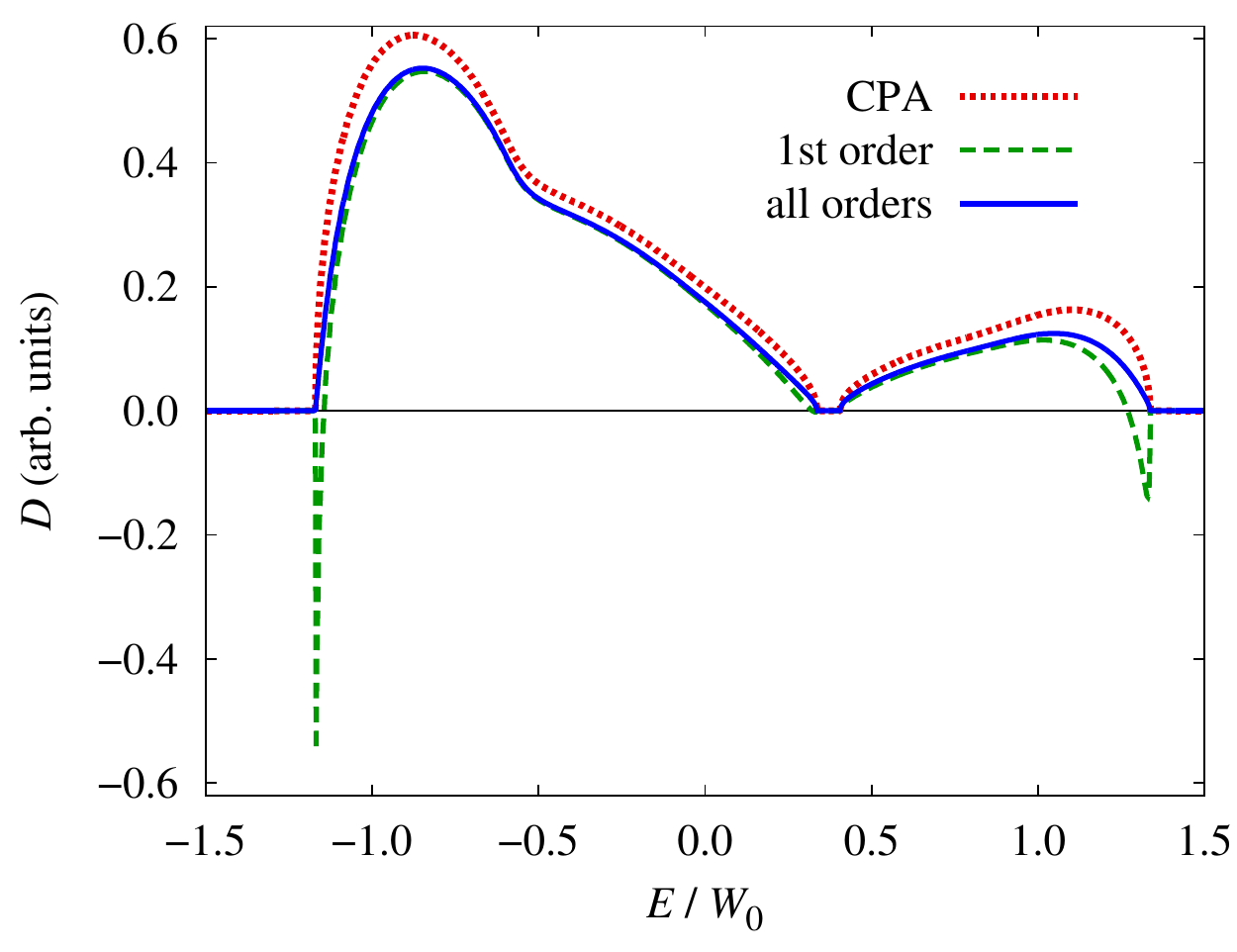}\hspace{10pt}\includegraphics[width=8.5cm]{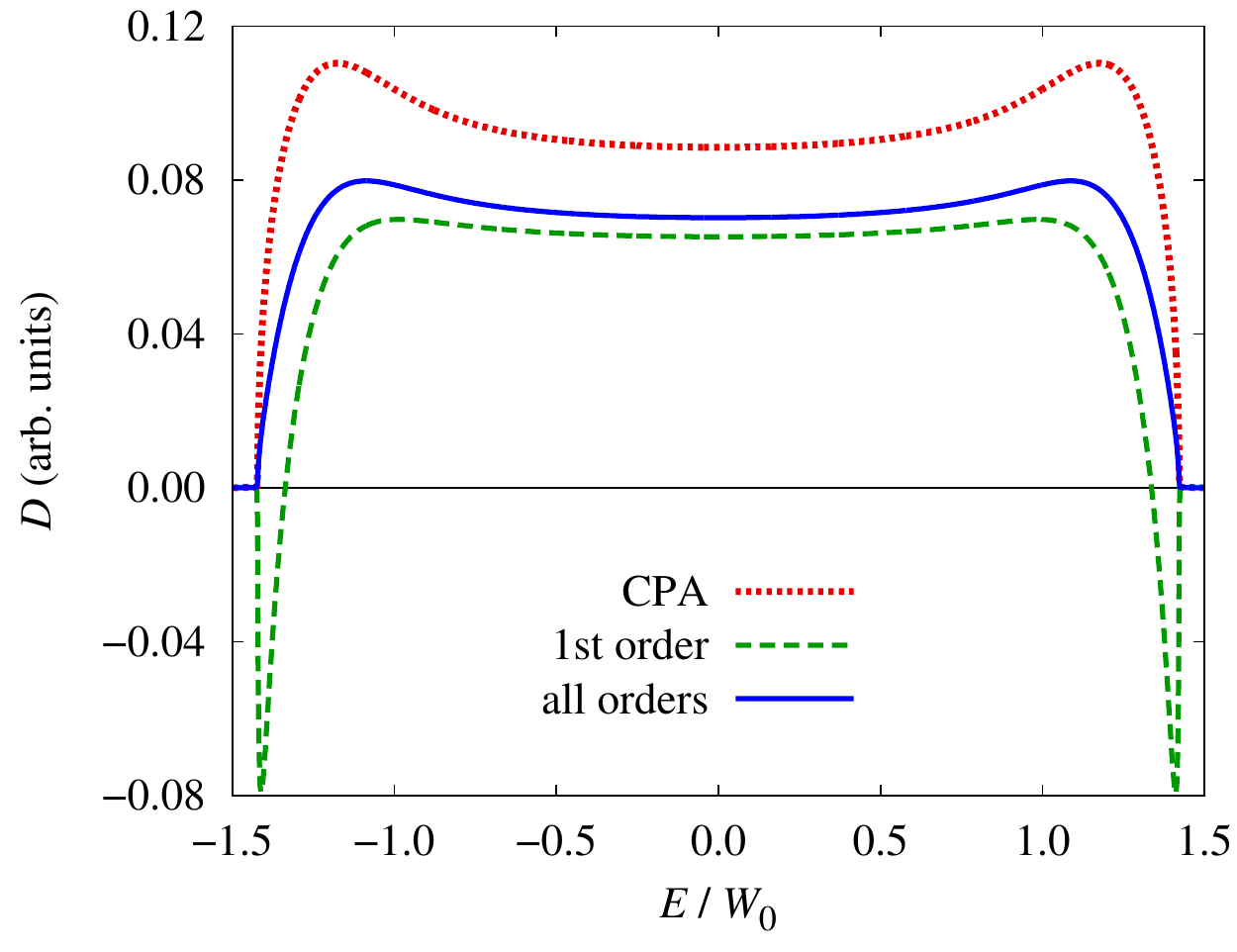}
\caption{(Color online) Diffusion constant for a binary alloy with component concentrations 30\% and 70\%, and with the difference between the local potentials $\Delta V=W_0$, where $2W_0$ is the band with of the lattice without disorder (left panel). Right panel shows analogous calculation for a random potential corresponding to the box distribution of width $2W_0$. Plotted are the values of the diffusion constant for the CPA, Eq.~\eqref{eq:DC-CPA} (dotted, red), CPA with the first-order vertex correction (dashed, green), and the full solution of Eq.~\eqref{eq:MFDC-correction} (solid, blue) on a simple cubic lattice.\label{fig:D_bin_box}}
\end{figure*}
The expression for the static diffusion constant in Eq.~\eqref{eq:MFDC-correction} is similar to the expression for vertex corrections to the mean-field conductivity due to maximally crossed diagrams derived in Ref.~\onlinecite{Pokorny:2013aa}. Although they both lead to a non-negative result and correctly describe weak localization in one- and two-dimensional systems, they are strictly justified only in the weak-scattering limit where the non-local corrections are negligible compared to the CPA self-energy. The diffusion constant from Eq.~\eqref{eq:MFDC-correction} is nevertheless attractive in that it offers an easily accessible  qualitative assessment of the impact of vertex corrections to the mean-field, Drude result without the necessity to go beyond the Soven equation for the local self-energy. Moreover, this result can easily be extended to multi-orbital models and used in realistic calculations.         

\begin{figure}
\includegraphics[width=8.5cm]{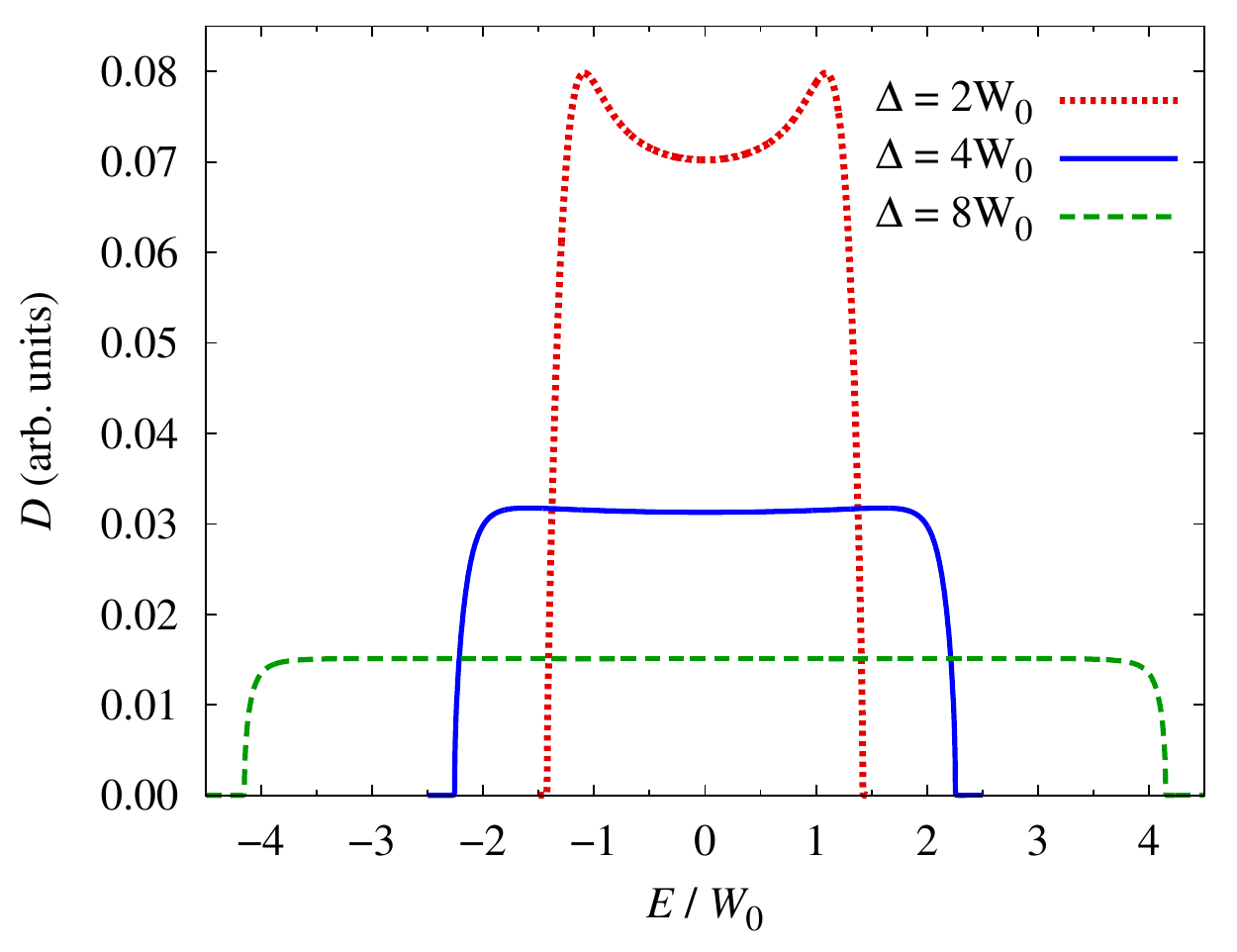}
\caption{(Color online) Diffusion constant from Eq.~\eqref{eq:MFDC-correction} for various box distributions of width $2W_0, 4W_{0}$, and $8W_{0}$ on a simple cubic lattice. The diffusion constant decreases with increasing disorder, but it does not reach zero for finite disorder strengths. \label{fig:D_box_var}}
\end{figure}

As an example, we calculated the static diffusion constant for a three-dimensional cubic lattice with the random potential corresponding to a binary alloy, and with a potential given by a box distribution. We compared the CPA result, Eq.~\eqref{eq:DC-CPA}, with a first-order vertex correction from the denominator on right-hand side of in Eq.~\eqref{eq:MFDC-correction}, and the full solution of Eq.~\eqref{eq:MFDC-correction}. The results are plotted in Fig.~\ref{fig:D_bin_box}. We can see that the values of the diffusion constant do not differ much deep inside the conduction bands, but the vertex corrections make the Drude result unstable near the band edges. Full solution of  Eq.~\eqref{eq:MFDC-correction} regularizes the correction so that the diffusion constant remains nonnegative. Neither of these solutions is, however, capable to describe the mobility edge or the Anderson localization transition. The diffusion constant decreases with increasing disorder but it never reaches zero for finite values of the disorder strength in this approximation, see Fig.~\ref{fig:D_box_var}. One needs a self-consistent approximation for the two-particle vertex in order to describe vanishing of diffusion at the Anderson localization transition in three and higher spatial dimensions.

 \section{Conclusions}
\label{Conclusions} 
 
Charge diffusion is a macroscopic phenomenon characterized by a specific form of the low-energy limit of the density-density response function. This function displays a diffusion pole where the long-range spatial fluctuations are controlled by a diffusion constant. Unfortunately, the density response function and the diffusion constant are not elementary objects of the microscopic theory of quantum diffusion. They are composite objects represented by the sum over the fermionic momenta of the two-particle Green function. This fact makes the full description of quantum diffusion from first, quantum-mechanical principles a demanding task that is very difficult to accomplish without inconsistencies and ad hoc or unjustified steps. The objective of this paper was to derive an exact representation of the diffusion pole and the diffusion constant in a consistent way from the elementary objects of the diagrammatic perturbation theory.    

A ubiquitous severe problem of the perturbation theory of non-interacting quantum random systems is the inability to obey the Ward identity when we go beyond the local, mean-field approximation. Whatever approach we choose, either an approximation for the self-energy (one-particle approach), or an approximation for the irreducible vertex (two-particle approach), the Ward identity cannot be guaranteed and is only used as a consistency check. In this paper, we made a substantial step forward and clarified the way  the two-particle irreducible vertices from the perturbation theory should be treated in order to restore the full dynamical Ward identity and all macroscopic conservation laws of measurable quantities. We used the Bethe-Salpeter equation with the irreducible vertex complying with the Ward identity and derived an exact low-energy singular asymptotics of the full two-particle vertex. We then represented the diffusion pole and the diffusion constant in the low-energy asymptotics of the electron-hole correlation function via the functions resulting from the perturbation expansion, that is, the self-energy and the two-particle irreducible vertex. It follows from our analysis that the singular structure of the two-particle vertex is more complex than that of the electron-hole correlation function. Consequently, the diffusion constant cannot be directly pulled into the perturbation theory as a parameter of a two-particle function to be self-consistently determined.          
   
The diffusion constant remains a parameter set apart of the perturbation theory. It contains only a reduced information from the two-particle vertex. It is represented as a matrix element of the full two-particle vertex via a Kubo-like formula. It follows from the exact representation derived here that any local approximation to the two-particle vertex irreducible in the electron-hole scattering channel results in the mean-field diffusion free of vertex corrections. Only non-local approximations with momentum-dependent irreducible vertices produce corrections to the Drude term. The contribution from the maximally crossed diagrams was explicitly calculated assuming dominance of the low-energy singularity in the two-particle vertex. We obtained a closed form of the static diffusion constant containing the leading vertex corrections and describing the weak localization that can easily be generalized to multi-orbital models and can be used in realistic calculations. Since the diffusion constant is not an integral part of the perturbation theory, finding a criterion of Anderson localization in terms of the irreducible vertex functions is still an open problem.  It also remains unclear what is the minimal approximation being able to describe qualitatively correctly the Anderson localization transition. A framework for addressing these questions was set in this paper.

 \section*{Acknowledgments}
Research on this problem was supported by Grant No. 15-14259S of the Czech Science Foundation. VJ thanks the Fulbright Commission for financing his stay at Louisiana State University where most of the research was performed. Access to computing and storage facilities owned by parties and projects contributing to the National Grid Infrastructure MetaCentrum, provided under the program LM2010005, is appreciated.

\end{document}